
 %
 %
\input harvmac.tex  
 %
\catcode`@=11
\def\rlx{\relax\leavevmode}                  
 %
 %
 %
\font\tenmib=cmmib10
\font\sevenmib=cmmib10 at 7pt 
\font\fivemib=cmmib10 at 5pt  
\font\tenbsy=cmbsy10
\font\sevenbsy=cmbsy10 at 7pt 
\font\fivebsy=cmbsy10 at 5pt  
\def\BMfont{\textfont0\tenbf \scriptfont0\sevenbf
                              \scriptscriptfont0\fivebf
            \textfont1\tenmib \scriptfont1\sevenmib
                               \scriptscriptfont1\fivemib
            \textfont2\tenbsy \scriptfont2\sevenbsy
                               \scriptscriptfont2\fivebsy}
\def\BM#1{\rlx\ifmmode\mathchoice
                      {\hbox{$\BMfont#1$}}
                      {\hbox{$\BMfont#1$}}
                      {\hbox{$\scriptstyle\BMfont#1$}}
                      {\hbox{$\scriptscriptstyle\BMfont#1$}}
                 \else{$\BMfont#1$}\fi}
 %
 %
 %
 %
\def\inbar{\vrule height1.5ex width.4pt depth0pt}
\def\sinbar{\vrule height1ex width.35pt depth0pt}
\def\ssinbar{\vrule height.7ex width.3pt depth0pt}
\font\cmss=cmss10
\font\cmsss=cmss10 at 7pt
\def\ZZ{\rlx\leavevmode
             \ifmmode\mathchoice
                    {\hbox{\cmss Z\kern-.4em Z}}
                    {\hbox{\cmss Z\kern-.4em Z}}
                    {\lower.9pt\hbox{\cmsss Z\kern-.36em Z}}
                    {\lower1.2pt\hbox{\cmsss Z\kern-.36em Z}}
               \else{\cmss Z\kern-.4em Z}\fi}
\def\Ik{\rlx{\rm I\kern-.18em k}}  
\def\IC{\rlx\leavevmode
             \ifmmode\mathchoice
                    {\hbox{\kern.33em\inbar\kern-.3em{\rm C}}}
                    {\hbox{\kern.33em\inbar\kern-.3em{\rm C}}}
                    {\hbox{\kern.28em\sinbar\kern-.25em{\sevenrm C}}}
                    {\hbox{\kern.25em\ssinbar\kern-.22em{\fiverm C}}}
             \else{\hbox{\kern.3em\inbar\kern-.3em{\rm C}}}\fi}
\def\IP{\rlx{\rm I\kern-.18em P}}
\def\IR{\rlx{\rm I\kern-.18em R}}
\def\Ione{\rlx{\rm 1\kern-2.7pt l}}
 %
 %

 %

\def\intem#1{\par\leavevmode%
              \llap{\hbox to\parindent{\hss{#1}\hfill~}}\ignorespaces}
 %


 %
\newskip\humongous \humongous=0pt plus 1000pt minus 1000pt  
\def\caja{\mathsurround=0pt}
\newif\ifdtup
 %
\def\eqalign#1{\,\vcenter{\openup2\jot \caja
     \ialign{\strut \hfil$\displaystyle{##}$&$
      \displaystyle{{}##}$\hfil\crcr#1\crcr}}\,}
 %
\def\twoeqsalign#1{\,\vcenter{\openup2\jot \caja
     \ialign{\strut \hfil$\displaystyle{##}$&$
      \displaystyle{{}##}$\hfil&\hfill$\displaystyle{##}$&$
       \displaystyle{{}##}$\hfil\crcr#1\crcr}}\,}
 %

 %

 %

 %

 %

 %
 %
 %
 %
   \let\SS=\S       
\def\,{\hskip1.5pt}           
 %
\let\a=\alpha
\let\b=\beta
\let\c=\chi
\let\d=\delta       \let\vd=\partial             \let\D=\Delta
\let\e=\epsilon     \let\ve=\varepsilon
\let\f=\phi                       
\let\g=\gamma                                    \let\G=\Gamma

\let\l=\lambda                                   

\let\n=\nu
\let\p=\pi          \let\vp=\varpi               
\let\q=\theta                   \let\Q=\Theta
\let\r=\rho         
\let\s=\sigma                   \let\S=\Sigma

\let\w=\omega                                    \let\W=\Omega
\let\x=\xi

 %
 %
\def\Box{\sqcap\llap{$\sqcup$}}
\def\lapp{\lower.4ex\hbox{\rlap{$\sim$}} \raise.4ex\hbox{$<$}}
\def\gapp{\lower.4ex\hbox{\rlap{$\sim$}} \raise.4ex\hbox{$>$}}
\def\con{\ifmmode\raise.1ex\hbox{\bf*}
          \else\raise.1ex\hbox{\bf*}\fi}
\def\bo{{\raise.15ex\hbox{\large$\Box\kern-.39em$}}}

\def\Ree{\mathop{\Re e}}

\def\dual{\relax\leavevmode\lower.9ex\hbox{\titlerms*}}
\def\define{\buildrel\rm def\over =}

\let\8=\otimes
 %
 %
 %
 %

\let\2=\underline
\let\ha=\widehat
\let\Tw=\widetilde
 %
\def\dt#1{{\buildrel{\smash{\lower1pt\hbox{.}}}\over{#1}}}

\font\eightrm=cmr8
\def\6(#1){\relax\leavevmode\hbox{\eightrm(}#1\hbox{\eightrm)}}
\def\0#1{\relax\ifmmode\mathaccent"7017{#1}     
                \else\accent23#1\relax\fi}      
\def\7#1#2{{\mathop{\null#2}\limits^{#1}}}      
\def\5#1#2{{\mathop{\null#2}\limits_{#1}}}      
 %

 %

 %

 %

 %
\newbox\t@b@x
\def\rightarrowfill{$\m@th \mathord- \mkern-6mu
     \cleaders\hbox{$\mkern-2mu \mathord- \mkern-2mu$}\hfill
      \mkern-6mu \mathord\rightarrow$}
\def\tooo#1{\setbox\t@b@x=\hbox{$\scriptstyle#1$}%
             \mathrel{\mathop{\hbox to\wd\t@b@x{\rightarrowfill}}%
              \limits^{#1}}\,}
\def\leftarrowfill{$\m@th \mathord\leftarrow \mkern-6mu
     \cleaders\hbox{$\mkern-2mu \mathord- \mkern-2mu$}\hfill
      \mkern-6mu \mathord-$}
\def\froo#1{\setbox\t@b@x=\hbox{$\scriptstyle#1$}%
             \mathrel{\mathop{\hbox to\wd\t@b@x{\leftarrowfill}}%
              \limits^{#1}}\,}
 %
\def\frac#1#2{{#1\over#2}}
\def\frc#1#2{\relax\ifmmode{\textstyle{#1\over#2}} 
                    \else$#1\over#2$\fi}           
 %
\def\Claim#1#2#3{\bigskip\begingroup%
                  \xdef #1{\secsym\the\meqno}%
                   \writedef{#1\leftbracket#1}%
                    \global\advance\meqno by1\wrlabeL#1%
                     \noindent{\bf#2}\,#1{}\,:~\sl#3\vskip1mm\endgroup}

\def\QED{\rlx\hfill$\Box$\kern-7pt\raise3pt\hbox{$\surd$}\bigskip}
 %
 %

 %
\def\muthstrut{\vphantom1}
\def\mutrix#1{\null\,\vcenter{\normalbaselines\m@th
        \ialign{\hfil$##$\hfil&&~\hfil$##$\hfill\crcr
            \muthstrut\crcr\noalign{\kern-\baselineskip}
            #1\crcr\muthstrut\crcr\noalign{\kern-\baselineskip}}}\,}

 %
\def\YT#1#2{\vcenter{\hbox{\vbox{\baselineskip0pt\parskip=\medskipamount%
             \def\Box{$\sqcap\llap{$\sqcup$}$\kern-1.2pt}%
              \def\Z{\hfil\vskip-5.8pt}\lineskiplimit0pt\lineskip0pt%
               \setbox0=\hbox{#1}\hsize\wd0\parindent=0pt#2}\,}}}
\def\EU{\rlx\ifmmode \c_{{}_E} \else$\c_{{}_E}$\fi}
\def\TM{\rlx\ifmmode {\cal T_M} \else$\cal T_M$\fi}
\def\TW{\rlx\ifmmode {\cal T_W} \else$\cal T_W$\fi}
\def\CM{\rlx\ifmmode {\cal T\rlap{\bf*}\!\!_M}
             \else$\cal T\rlap{\bf*}\!\!_M$\fi}
\def\hm#1#2{\rlx\ifmmode H^{#1}({\cal M},{#2})
                 \else$H^{#1}({\cal M},{#2})$\fi}
\def\CP#1{\rlx\ifmmode\IP^{#1}\else\IP$^{#1}$\fi}
\def\cP#1{\rlx\ifmmode\IC{\rm P}^{#1}\else$\IC{\rm P}^{#1}$\fi}

\def\sll#1{\rlx\rlap{\,\raise1pt\hbox{/}}{#1}}
\def\Sll#1{\rlx\rlap{\,\kern.6pt\raise1pt\hbox{/}}{#1}\kern-.6pt}

\let\SSS=\scriptstyle

 %
 %
\def\ie{\hbox{\it i.e.}}        
\def\CFT{conformal field theory}

\def\CY{Calabi-\kern-.2em Yau}
\def\LGO{Landau-Ginzburg orbifold}
\def\3{\ifmmode\ldots\else$\ldots$\fi}
\def\Z{\hfil\break\rlx\hbox{}\quad}
\def\3{\ifmmode\ldots\else$\ldots$\fi}
\def\?{d\kern-.3em\raise.64ex\hbox{-}}           
\def\9{\raise.43ex\hbox{-}\kern-.37em D}         
\def\ping{\nobreak\par\centerline{---$\circ$---}\goodbreak\bigskip}
 %
 %

 %

 %

\def\NP#1{{\it Nucl.\,Phys.\,}{\bf#1\,}}
\def\PL#1{{\it Phys.\,Lett.\,}{\bf#1\,}}

\def\CMP#1{{\it Commun.\,Math.\,Phys.\,}{\bf#1\,}}

 %
 %
 %
\baselineskip=13.0861pt plus2pt minus1pt
\parskip=\medskipamount
\let\ft=\foot
\noblackbox
\def\SaveTimber{\abovedisplayskip=1.5ex plus.3ex minus.5ex
                \belowdisplayskip=1.5ex plus.3ex minus.5ex
                \abovedisplayshortskip=.2ex plus.2ex minus.4ex
                \belowdisplayshortskip=1.5ex plus.2ex minus.4ex
                \baselineskip=12pt plus1pt minus.5pt
 \parskip=\smallskipamount
 \def\ft##1{\unskip\,\begingroup\footskip9pt plus1pt minus1pt\setbox%
             \strutbox=\hbox{\vrule height6pt depth4.5pt width0pt}%
              \global\advance\ftno by1\footnote{$^{\the\ftno)}$}{##1}%
               \endgroup}
 \def\listrefs{\footatend\vfill\immediate\closeout\rfile%
                \writestoppt\baselineskip=10pt%
                 \centerline{{\bf References}}%
                  \bigskip{\frenchspacing\parindent=20pt\escapechar=` %
                   \rightskip=0pt plus4em\spaceskip=.3333em%
                    \input refs.tmp\vfill\eject}\nonfrenchspacing}}
 %
\def\Afour{\ifx\answ\bigans
            \hsize=16.5truecm\vsize=24.7truecm
             \else
              \hsize=24.7truecm\vsize=16.5truecm
               \fi}
\catcode`@=12
 %
 %
\def\rd{{\rm d}}
\def\\{\hfill\break}

\def\cp#1#2{\ifmmode{\IP_{#1}^{#2}}\else$\IP_{#1}^{#2}$\fi}
\def\prp{positive real part}


\def\cropen#1{\crcr\noalign{\vskip #1}}
 %
 %
\Title{\vbox{\baselineskip12pt \hbox{HUPAPP-93/6}
                                 \hbox{IASSNS-HEP-93/80}
                                  \hbox{UTTG-27-93}}}
      {\vbox{\centerline{\titlerm On Periods for String
                                  Compactifications}}}
\centerline{\titlerms\hfil
            P.~Berglund$^1$,\hfil
            E.~Derrick$^2$,\hfil
            T.~H\"ubsch$^3$\footnote{$^{\spadesuit}$}
                      {On leave from the Institute ``Ru{\?}er
                         Bo\v{s}kovi\'{c}'', Zagreb, Croatia.}\hfil
            and\hfil
            D.~Jan\v{c}i\'{c}$^2$
}
\vskip 5mm
\line{\vtop{\baselineskip=11pt\hsize = 50mm
       \centerline{$^1$\it School of Natural Science}
       \centerline{\it Institute for Advanced Study}
       \centerline{\it Olden Lane}
       \centerline{\it Princeton NJ 08540}}\hfill
      \vtop{\baselineskip=11pt\hsize = 40mm
       \centerline{$^2$\it Theory Group}
       \centerline{\it Department of Physics}
       \centerline{\it University of Texas}
       \centerline{\it Austin, TX 78712}}\hfill
      \vtop{\baselineskip=11pt\hsize = 40mm
       \centerline{$^3$\it Department of Physics}
       \centerline{\it Howard University}
       \centerline{\it Washington DC 20059} }}
\vfill

\centerline{ABSTRACT}\vskip10mm
\vbox{\baselineskip=12pt\noindent
Motivated by recent developments in the computation of periods for string
compactifications with $c=9$, we develop a complementary method which also
produces a convenient basis for related calculations. The models are
realized as \CY\ hypersurfaces in weighted projective spaces of dimension
four or as Landau-Ginzburg vacua. The calculation reproduces known results
and also allows a treatment of \LGO{s} with more than five fields.}

\Date{November 1993}
\footline{\hss\tenrm--\,\folio\,--\hss}
\noblackbox
 %
\lref\rAS{A.~Strominger: \CMP{133} (1990) 163.}

\lref\rCd{P.~Candelas and X.~C.~de~la~Ossa: \NP{B355} (1991) 455.}

\lref\rCdGP{P.~Candelas, X.~ de la Ossa, P.~Green and L.~Parkes:
       \NP{B359} (1991) 21.}

\lref\rArnold{V.I.~Arnold: in {\it Proceedings of the International
       Congress of Mathematicians}, p.19, Vancouver, 1974,  {\it
       Singularity Theory} ({\it London Math.\ Soc.\ Lecture Note
       Series}~53, Cambridge University Press, Cambridge, 1981)\semi
      {V.I.~Arnold, S.M.~Gusein-Zade and A.N.~Varchenko}: {\it
       Singularities of Differentiable Maps} (Birkh\"auser, Boston,
       1985).}

\lref\rCVTaT{S.~Cecotti and C.~Vafa: Topological Anti-Topological
       Fusion. \NP{B367} (1991) 359--461.}

\lref\rMaxSkI{M.~Kreuzer and H.~Skarke: \NP{B388} (1992) 113.}

\lref\rMaxSkII{M.~Kreuzer and H.~Skarke: ``On the Classification
       of Quasihomogenous Functions'', CERN preprint CERN-TH-6373/92.}

\lref\rAR{A.~Klemm and R.~Schimmrigk: ``Landau-Ginzburg Vacua'',
       CERN preprint CERN-TH-6459/92, Universit\"{a}t Heidelberg report
       HD-THEP-92-13.}

\lref\rBH{P.~Berglund and T.~H\"ubsch: \NP{B393} (1993) 377, also in
       {\it Essays on Mirror Manifolds}, p.388, ed.\ S.-T.~Yau,
       (Intl.\ Press, Hong Kong, 1992).}

\lref\rCYCI{T.~H\"ubsch: \CMP{108} (1987) 291.}

\lref\rGH{P.~Green and T.~H\"ubsch: \CMP{109} (1987) 99.}

\lref\rCDLS{P.~Candelas, A.M.~Dale, C.~A.~L\"utken and R.~Schimmrigk:\Z
       \NP{B298} (1988) 493.}

\lref\rBeast{T.~H\"ubsch: {\it \CY\ Manifolds---A Bestiary for
       Physicists}\Z (World Scientific, Singapore, 1992).}

\lref\rCLS{P.~Candelas, M.~Lynker and R.~Schimmrigk: \NP{B341} (1990) 383.}

\lref\rLS{M.~Lynker and R.~Schimmrigk: \PL{249B} (1990) 237.}

\lref\rCfC{P.~Berglund and T.~H\"ubsch: ``Couplings for
       Compactification'', {\it Howard University preprint} HUPAPP-93/2,
       \NP{B~}(in press).}

\lref\rABG{M.~Atiyah, R.~Bott and L.~G{\aa}rding:
      {\it Acta Math.\,\bf131} (1973) 145.}

\lref\rPhilip{P.~Candelas: \NP{B298} (1988) 458.}

\lref\rErdelyi{A.~Erd\'elyi, F.~Oberhettinger, W.~Magnus and
       F.~G.~Tricomi: {\it Higher Transcendental Functions, 3 Vols.}
       (McGraw--Hill, New York, 1953).}

\lref\rGRY{B.~Greene, S.-S.~Roan and S.-T.~Yau: \CMP{142}(1991)245.}

\lref\rZ{P.~Candelas, E.~Derrick and L.~Parkes:
       ``Generalized Calabi-Yau Manifolds and the Mirror of a Rigid
       Manifold'', \NP{B407} (1993) 115.}

\lref\rGoryVII{P.~Berglund, P.~Candelas, X.~de la Ossa, A.~Font,
       T.~H\"ubsch, D.~Jan\v{c}i\'{c} and F.~Quevedo: Periods for \CY\ and
       Landau-Ginzburg Vacua, CERN preprint CERN-TH 6865/93.}

\lref\rRolling{P.~Candelas, P.S.~Green and T.~H\"ubsch:
      \NP{B330} (1990) 49.}

\lref\rDG{J.~Distler and B.R.~Greene: \NP{B309} (1988) 295.}

\lref\rMatter{P.~Berglund, T.~H\"ubsch and L.~Parkes: \CMP{148} (1992) 57.}

\lref\rVW{C.~Vafa and N.P.~Warner: \PL{218B} (1989) 377\semi
          E.Martinec: \PL{B217} (1989) 431.}

 %
%
\newsec{Introduction, Results and Summary}\seclab\sIRS\noindent
Until recently, the most general known superstring models with
worldsheet (2,2) supersymmetry were also the least amenable to calculation
and physics prediction. Such models are built on \CY\ spaces, $\cal M$,
and depend on their complex structure and (complexified) K\"ahler class:
the dynamics of matter fields, respectively the {\bf27}'s and {\bf27\con}'s
of $E_6$, is determined by the so-called special geometry of these two
sectors of the parameter space\penalty10000\ft{In addition,
a subset of matter
fields, being {\bf1}s of $E_6$, have no Yang-Mills interaction and
their dynamics is determined by another, rather less well understood
parameter space.} (for a comprehensive review, see~\rBeast \ and
references therein).
By a remarkable characteristic of special
geometry, both the kinetic and the coupling terms are determined from a
single object which is complex analytic over the parameter
space~\refs{\rRolling, \rAS, \rCd}. The dynamics of the {\bf27}'s is
governed by the functional dependence over the parameter space of the
holomorphic volume-form $\W$ on $\cal M$. It is convenient to express $\W$
as a vector over a basis of homology cycles in $\cal M$ with components
$\vp_j\define\oint_{\g^j}\W$, which are called periods.

At special subregions of the parameter space, the underlying
(2,2)-super\CFT\ simplifies sufficiently so as to enable
effective and accurate calculation of physical data. The generic models,
however, remain understood only in terms of their \CY\ geometry, and were
widely believed to admit in practice only perturbative calculations of
limited accuracy and unlimited difficulty.
 Certain non-renormalization theorems~\rDG\ do ensure the exactness of the
${\bf27}^3$ couplings for which a rather successful technology has
developed~\refs{\rPhilip, \rMatter, \rCfC};
using special geometry and the
`mirror map' (for cases where the mirror model {\it is} known~\rBH)
then allows the determination of a large portion of the dynamics of such
{\it particular} models.
 Finally, combining these facts in an essential way with the complex
analyticity over the parameter space has brought forth an approach
of hitherto unwitnessed calculational power. First developed on a
1-parameter example~\rCdGP, and now generalized to a wealth of large
families of examples~\rGoryVII, this is absolutely the most widely
applicable exact technique.

The results of this article are twofold.
 Firstly, following a suggestion in Ref.~\rGoryVII, (see also
Refs.~\refs{\rCdGP,\rZ}), we
develop a more direct calculation of periods, rather than threading through
analytic continuation and modular group action as in Refs.~\refs{\rCdGP,
\rGoryVII}. This more direct calculation is applicable to
\CY\ spaces, not necessarily of dimension three,
described as a single hypersurface in a weighted projective space.
More importantly, we are also able to handle string vacua
expressed as \LGO{s} with more than five fields.
The geometrical interpretation of these models is not always known.
Nevertheless the periods which we obtain are solutions to a Picard-Fuchs
equation whose existence is assured since the models are $(2,2)$ string
vacua.
 Second, as an additional benefit of this approach, a convenient and
rather simple homology basis is produced, which allows a simple and
intuitive description of periods and is also suitable for
topological--anti-topological fusion calculations~\rCVTaT.

The paper is organized as follows: in section~2 we describe the
technique by which periods are constructed. Section~3 is devoted to some
non-trivial examples, while discussions and conclusions are left for
section~4. Some technical details regarding our choice of contours
are presented in the appendix.

\newsec{The Construction}\seclab\sBasTyp\noindent
The \CY\ spaces studied herein will all be defined as complete
intersections of hypersurfaces in some (weighted) projective space, each
one defined as the zero-set of a defining polynomial~\refs{\rBeast, \rCYCI,
\rGH, \rCDLS, \rGRY}. In fact, we wish to consider families of
such spaces, parametrized by the coefficients in the defining polynomials.
This only parametrizes the complex structure moduli space. However,
using mirror symmetry, this gives us a handle on the K\"ahler moduli space
as well; starting with a mirror pair $({\cal M}, {\cal W})$, we study
the parameter space of complex structure deformations for ${\cal M}$
{\it and} ${\cal W}$. By the mirror symmetry conjecture this result
is the equivalent of the quantum corrected moduli space of the K\"ahler
class corresponding to ${\cal W}$ and ${\cal M}$ respectively.

Begin with the case of a family of hypersurfaces
${\cal M}_{\f}$ defined as the zero-set of a defining polynomial $P_\f$.
The periods are defined as
\eqn\ePerDef{ \vp_j(\f_\a)~~ \define ~~\int_{\g^j} \W(\f_\a)~, }
where $\W(\f_\a)$ is the nowhere vanishing holomorphic
3-form~\refs{\rPhilip, \rABG}
\eqn\eOmega{ \W(\f_\a)~~ \define
           ~~{\rm Res}_{{\cal M}_\f}\bigg[{(x\rd^4x)\over P_\f}\bigg]~, }
on the \CY\ 3-fold specified by the parameters $\f_\a$. The 4-differential
$(x\rd^4x)$ is the `natural' one: on a weighted projective $N$-space
 \cp{(k_0,\ldots,k_N)}{N}, we have
\eqn\eVolume{ (x\rd^Nx)~~ \define
             ~~{1\over (N{+}1)!} \e^{i_0 i_1\cdots i_N}
                k_{i_0} x_{i_0} \rd x_{i_1} \cdots \rd x_{i_N}~, }
where $k_i$ are the weights of the coordinates. That is, in
 \cp{(k_0,\ldots,k_N)}{N},
\eqn\eWProj{ (x_0,\ldots,x_N) \cong (\l^{k_0}x_0,\ldots,\l^{k_N}x_N)~,
             \qquad \l \in \IC^*~. }
The period~\ePerDef\ will here be calculated by choosing one of the
standard coordinate patches in \cp{(k_0,\ldots,k_4)}{4}, ${\cal U}_m$,
where $x_m\neq0$ so that $x_m=1$ by projectivity. There~\rGoryVII:
\eqn\eOM{ \vp^{(m)}_j(\f_\a)~~ = ~~C \int_{\G^j}
          {\prod_{i\neq m}\rd x_i\over \big[P_\f\big]_{x_m=1}}~, }
with $C$ a convenient prefactor. This is easily seen to apply
for \CY\ weighted hypersurfaces of arbitrary dimension.

We separate the polynomial into a reference polynomial $P_0$
independent of the moduli~$\f$, and a perturbative part $\D$
which does depend on the $\f$.
This amounts to choosing a reference point in moduli space,
and expanding around that point.
The basic idea here is to
expand $1/P_\f$ around $1/P_0$, and utilize the Laplace transform
\eqn\eLaplace{ {1\over P_\f}~ = ~
{1\over P_0 - \D} ~=~ \sum_{n=0}^\infty {\D^n \over P_0^{n+1} }
{}~=~ \sum_n \D^n \int_0^\infty ds {s^n \over n!} e^{-s P_0}
{}~,\qquad \Ree(P_0)>0~,}
which produces a ``small-$\f$''
expansion of the periods~\eOM.

With a choice of the poly-contours $\{\G^j\}$ discussed below for the
various types of polynomials, Eq.~\eOM\ may be considered a
definition of the periods.

We now discuss suitable contours for all types of reference polynomials
found in \refs{\rCfC,\rCLS}.  The polynomials are built out
of basic patterns labelled Fermat types, tadpoles and loops,
which we treat in turn.

\subsec{Fermat Models}\subseclab\sBPFWZ\noindent
Write the defining polynomial as
\eqn\eDefP{ P_\f~ = ~P_0~ - ~\sum_\a \f_\a M_\a~, }
and consider the case where
\eqn\ePOF{ P_0~ = ~\sum_{i=0}^N x_i^{a_i} }
is the reference polynomial of the Fermat type. The deformation terms,
\eqn\eMa{ M_\a~ = ~\prod_{i=0}^N x_i^{q_i^\a}~,\qquad \a=1,\ldots,M~,  }
are suitable monomials. Writing $d=\deg(P_\f)$, we have that, for
homogeneity,
\eqn\eCond{{\eqalign{           a_i k_i~&=~d~, \qquad \forall\,i~, \cr
                     \sum_i k_i q_i^\a ~&=~d~, \qquad \forall\,\a~.\cr}}}
For the hypersurface to be \CY, we also need
\eqn\eCondCY{ \sum_i k_i~ = ~d~. }

In the affine coordinate patch ${\cal U}_m$ where $x_m=1$,
the defining polynomial becomes
\eqn\eUmP{
   P_\f~~ = ~~1 + \sum_i\strut' x_i^{a_i}~ -
                  ~\sum_\a \f_\a \prod_i\strut' x_i^{q_i^\a}~, }
where the prime on the summation and product sign denotes that the
indicated summation and product skip $i=m$. In this patch, and
supressing the homology labels, the periods become
\eqn\eXXX{ \vp^{(m)}(\f_\a)~ = ~C \int_\G \rd^4 x
            \int_0^\infty \rd s~
\sum_{n=0}^\infty {s^n (\sum_\a \f_\a M_\a)^n \over n!}
e^{-sP_0}~,\qquad \Ree(P_0)>0~. }
The poly-contours $\G$ have to be chosen so that the
convergence criterion $\Ree(P_0)>0$ is obtained.
To this end, label the poly-contours $\G$ by the $N$-vector
$(\d_0,\ldots,\widehat\d_m,\ldots,\d_N)$, the caret denoting omission.
That is, the poly-contour $\G$ is a product of $N$ contours, the
$i$'th of which is V-shaped and connects the points
 $(\l^{k_i(\d_i+1)}\infty, 0, \l^{k_i\d_i}\infty)$ in the $x_i$ plane
with two straight lines called `spokes';
here $\l=e^{2\p i\over d}$, and $\d_i=0,1,\ldots,a_i{-}2$;
the $\d_i=a_i{-}1$ contour is easily seen to be the sum (concatenation)
of the others and is hence omitted as uninteresting.
The choice of these contours ensures that each $x_i^{a_i}>0$, whence the
integrals converge. It is amusing to note that this set of rather simple
choices often suffices.

Expanding ~\eUmP,
\eqn\eXXX{{\eqalign{ \vp^{(m)}_{(\d_0,\ldots,\widehat\d_m,\ldots,\d_N)}~
 &= ~C~ \sum_{n_\b=0}^\infty
       ~ \prod_\b{\f_\b^{~n_\b}\over n_\b !}
\int_0^\infty\!\!\rd s~s^{\|n\|}\> e^{{-}s} I_\d(s) ~, \cr
   I_\d(s)~
 &\define~\prod_i\strut' \int_{V(\d_i)}\!\!\rd x_i\>
           e^{{-}s\>x_i^{a_i}}~ x_i^{q_i{\cdot}n } ~, \cr}}}
where $\int_{V(\d_i)}\rd x_i$ denotes the integral in the
$x_i$-plane, along the V-shaped contour connecting the points
$(\l^{k_i(\d_i+1)}\infty, 0, \l^{k_i\d_i}\infty)$, and
\eqn\eXXX{ \|n\|~       \define ~\sum_\a n_\a~,\qquad
           q_i{\cdot}n~ \define ~\sum_\a q_i^\a n_\a~. }
Next, we use
\eqn\eXXX{ \int_{V(\d_i)}\!\!\rd\x\>\x^\n\,e^{{-}s\,\x^{a_i}}~
  = ~\l^{k_i\d_i(\n+1)} \big(1-\l^{k_i(\n{+}1)}\big)\, {k_i\over d}\,
    {\G\big({k_i(\n+1)\over d}\big)\over s^{k_i(\n+1)/d}}~. }
Therefore
\eqn\eXXX{{\eqalign{ \vp^{(m)}_{(\d_0,\ldots,\widehat\d_m,\ldots,\d_N)}~
 &= ~{C\over d^N}~ \sum_{n_\b=0}^\infty
  ~\prod_\b{\f_\b^{~n_\b}\over n_\b !}
  \left[\int_0^\infty\!\!\rd s~e^{{-}s}\>
        s^{\big[{k_m\over d}\big(1{+}q_i{\cdot}n\big){-}1\big]}\right]\cr
 &\qquad\quad\times \prod_i\strut' k_i~
   \l^{k_i\d_i(1{+}q_i{\cdot}n)}
    \big( 1{-}\l^{k_i(1{+}q_i{\cdot}n)} \big)
     \G\Big({k_i\over d}\big(1{+}q_i{\cdot}n\big)\Big)~,\cr}}}
and we find that
\eqn\eExVPF{{\eqalign{\vp^{(m)}_{(\d_0,\ldots,\widehat\d_m,\ldots,\d_N)}~
 &~={C\prod_j k_j \over k_m d^N}~\Big[ \prod_i\strut' \l^{k_i\d_i}\Big]~
     \sum_\b \prod_\b{\Big( \l^{(\S_j\!\!'\d_j k_j q_i^\b)}
                             \f_\b\Big)^{n_\b} \over n_\b !}        \cr
&\qquad\times\prod_i \G\Big({k_i\over d}\big(1{+}q_i{\cdot}n\big) \Big)~
    \prod_j\strut' \left( 1{-}\l^{k_j(1{+}q_i{\cdot}n)} \right)~.   \cr}}}

There are $a_i{-}1$ choices of $\d_i$ for each $i=0,\ldots,N$, $i\neq m$,
yielding altogether $\prod_i\!\!\strut'(a_i{-}1)$ possible poly-contours.
This then is the upper limit on the number of distinct periods~\eExVPF. In
the worked examples, this turns out to be bigger than the total number of
periods and, in fact, all $b_3=2(b_{2,1}{+}1)$ periods may be represented
in this way. This construction of cycles
from spokes is further detailed in the appendix.
 \ping

Consider, on the other hand, the integral over the arc at infinity
subtended between the two spokes of the above V-shaped contour.
On the arc, take $x=r e^{i\q}$ and let $\q$ run from ${2\p\d\over a}$ to
${2\p(\d{+}1)\over a}$. Then
\eqn\eXXX{
I = \lim_{r\to\infty}
  \int \rd x \ x^\s \ e^{-s x^a}~~
  = ~~\lim_{r\to\infty}
   i r^{\s+1} \, \int_{2\p \d\over a}^{2\pi(\d+1)\over a}
          \rd \q \ e^{i(\s+1)\q -s r^a e^{ia\q}}~. }
Since $r^a$ is large, we expand in a power series, do the integral,
and find a confluent hypergeometric function:
\eqn\eXXX{{\eqalign{
I &= \lim_{r\to\infty} \sum_{n=0}^\infty
   {i (-s)^n\, r^{\s{+}1{+}an}\over n!}\>
    \int_{2\p\d\over a}^{2\p(\d+1)\over a}\rd\q~ e^{i\q(\s+1+an)} \cr
 &= e^{2\p i\d{\s+1\over a}} \big(e^{2\p i {\s+1\over a}}{-}1\big)
    \lim_{r\to\infty} \sum_{n=0}^\infty
                           {(-s)^n\, r^{\s+1+an}\over (\s+1+an) n!} \cr
 &= e^{2\p i \d{\s+1\over a}} \big(e^{2\p i {\s+1\over a}}{-}1\big)
    \lim_{r\to\infty} \sum_{n=0}^\infty
      {r^{\s+1}\over a} {(-s r^a)^n \over n!}
 {\G\big(n+{\s+1\over a}\big) \over \G\big(n+1+{\s+1\over a}\big) } \cr
 &= e^{2\p i \d{\s+1\over a}} \big(e^{2\p i {\s+1\over a}}{-}1\big)
    \lim_{r\to\infty} {r^{\s+1} \over \s+1}
  {}_1F_1\left({\s+1\over a},{\s+1\over a}+1;-sr^a\right)~.\cr}}}

Now the limit from Ref.~\rErdelyi, p.278, noting that
the argument is going to $-\infty$, finally yields
\eqn\eXXX{{\eqalign{
I &= e^{2\p i \d{\s+1\over a}} \big(e^{2\p i {\s+1\over a}}-1 \big)
     \lim_{r\to\infty}{r^{\s+1} \over \s+1} \G\big({\s+1\over a}+1\big)\,
                                     \big(sr^a\big)^{-{\s+1\over a}}\cr
  &={1 \over a}
     e^{2\p i \d{\s+1\over a}} \big(e^{2\p i {\s+1\over a}}-1 \big)
      \G\big({\s+1\over a}\big)\, s^{-{\s+1\over a}}~,\cr}}}
which is exactly the same as the contribution from the V-shaped contour
used above, except for a minus sign. Indeed, owing to the analyticity of
the integrand in the wedge bounded by the V-shaped contour and the arc at
infinity, these two contours may be deformed into each other at will.

\subsec{Tadpole Models}\subseclab\sTadpoles\noindent
Following the polynomial classification methods of
Arnold~\refs{\rBH, \rCLS, \rArnold, \rMaxSkII}, consider next the simplest
of `tadpole' type polynomials,
\eqn\eRefPT{ P_0^{(\r)}~ = ~x_0^{a_0} + x_0 x_1^{a_1}~. }
In the course of calculating the periods one encounters integrals of the
form
\eqn\eTadGenPer{
 \int_{\g_0} \rd x_0 \int_{\g_1} \rd x_1~
   e^{-s \left( x_0^{a_0} + x_0 x_1^{a_1} \right)}\>
   x_0^{\s_0}\> x_1^{\s_1} ~, }
wherein we must choose the contour $\g_0 \times \g_1$ to ensure that the
tadpole has \prp. We will do this by requiring each term to have a \prp.
To ensure this for the `Fermat' term, $x_0^{a_0}$, we restrict $x_0$ to
be on the spokes
\eqn\eXXX{ x_0 = \x_0 \, e^{2\p i {r \over a_0}}~,\qquad
           r \in \ZZ~,\quad 0\leq\x_0<\infty~. }
We can then choose a contour $\g_0$ where $x_0$ follows
 $x_0 = \x_0 \, e^{2\p i {\d_0+1 \over a_0}}$
from $\infty$ to 0,  and then
 $x_0 = \x_0 \, e^{2\p i {\d_0 \over a_0}}$
from 0 to $\infty$. Label the contour by $\d_0$, for which the values
$0,\ldots,a_0{-}2$ describe independent paths.

To ensure that the term $x_0 x_1^{a_1}$ has a \prp , we require that
\eqn\eXXX{ \arg(x_1^{a_1})~ = ~-2\p{( \d_0 + {1\over 2})\over a_0}~. }
This means that the argument of $x_0 x_1^{a_1}$ will be either
 $-{\p\over a_0}$ or ${\p\over a_0}$,  and so as long as $a_0>2$,
this has a \prp. In the special case when $a_0=2$, we may change variables
to a Fermat-type polynomial (see below).
This means that $x_1$ lies on spokes defined by
\eqn\eXXX{ x_1~ = ~\x_1\, e^{2\p i\big( {\SSS s\over a_1}
                             {-}{\d_0+{1\over2}\over a_1 a_0}\big)}~,
           \qquad s\in \ZZ~,\quad 0\leq\x_1<\infty~. }
Taking adjacent spokes, we let $x_1$ follow the spoke with
 $s=\d_1+1$ from $\infty$ to 0, and the spoke $s=\d_1$ from 0 to $\infty$,
and we may choose $\d_1$ from $0,\ldots,a_1{-}2$.

We may now do the integral \eTadGenPer
\eqn\eTadPer{
{\eqalign{
I &= \int_{\d_0 \times \d_1} \rd x_0\, \rd x_1~
   e^{-s \left( x_0^{a_0} + x_0 x_1^{a_1} \right)}\,
   x_0^{\s_0} x_1^{\s_1} \cr
  &= {1\over a_0 a_1}
  e^{2\p i\left(\d_1({\s_1 +1\over a_1})
                +\d_0({\s_0 +1\over a_0}-{\s_1 +1\over a_1 a_0}) \right)}
  \left[1{-}e^{2\p i({\s_1 +1\over a_1})}\right]
  \left[1{-}e^{2\p i({\s_0 +1\over a_0}-{\s_1{+}1\over a_1 a_0})}\right]\cr
  &\qquad\qquad\times\G\Big({\s_1{+}1\over a_1}\Big)
  \G\Big({\s_0{+}1\over a_0}{-}{\s_1{+}1\over a_1 a_0}\Big)
  s^{-({\s_0 +1\over a_0}+{(\s_1 +1)(a_0-1)\over a_1 a_0})} ~.\cr}}}

The extension to longer tadpole polynomials is straightforward.
If the polynomial contains the third term, $x_1\,x_2^{a_2}$,
then we insist that $x_2$ lie along the spokes
\eqn\eXXX{ x_2 = \x_2\, e^{2\p i\big( {\SSS t\over a_2}
                           {-}{\d_1+{1\over2}\over a_2 a_1}
                           {+}{\d_0+{1\over2}\over a_2 a_1 a_0}\big)}~,
           \quad t\in\ZZ~,\quad 0\leq\x_2<\infty~. }
For adjacent spokes, take $t=\d_2+1$ on the spoke towards the origin and
$t=\d_2$ on the way out. The corresponding integral is then
obtained along the lines of~\eTadPer.
The diligent Reader will have no problem in iterating this procedure to
obtain the integrals analogous to~\eTadPer.

\subsec{Loop Models}\subseclab\sLoops\noindent
Next, consider the simplest loop-type polynomial:
\eqn\eLoop{ P_0^{(O)} = x_0^{a_0} x_1 + x_0 x_1^{a_1}~. }
We assume that $x_0$ and $x_1$ will follow spokes so that for each
$j=0,1$,
\eqn\eXXX{ {\eqalign{
 \arg x_{j}^{a_{j}} = 2\p\a_{j} & \qquad{\rm implying}\qquad
     x_{j} = \xi_{j} \, e^{2\p i{\a_{j} + m\over a_{j}}}   }}}
Choosing adjacent spokes for each ($x_0$ comes in
from $\infty$ on $\d_0+1$, out on $\d_0$,
$x_1$ comes in
from $\infty$ on $\d_1+1$, out on $\d_1$)
means that the terms of the polynomial lie along the spokes
\eqn\eXXX{ {\eqalign{
 \arg (x_{j}^{a_{j}} x_{1-j}) =
2\p\left(\a_{j}+{\a_{1-j}+\d_{1-j}+1\over a_{1-j}}\right)
   \qquad &{\rm or}\qquad2\p\left(\a_{j}+{\a_{1-j}+\d_{1-j}\over
      a_{1-j}}\right)\cr       }}}
We centre these on the positive real axis
by solving the equations
\eqn\eloopmtx{\eqalign{
\a_{j}+{\a_{1-j}+\d_{1-j}+{1\over 2}\over a_{1-j}} &=0\cr
}}
which has the solutions
\eqn\eXXX{ {\eqalign{
\a_j&={\d_{j} + {1\over 2}-a_j\left(\d_{1-j}+{1\over 2}\right)
        \over a_{0} a_{1} -1}\cr    }}}
The integral now becomes
\eqn\eloopper{
{\eqalign{
I &= \int_{\d_0 \times \d_1} \rd x_0\, \rd x_1~
   e^{-s \left( x_0^{a_0} x_1 + x_0 x_1^{a_1} \right)}\,
   x_0^{\s_0}\> x_1^{\s_1} \cr
  &= {1\over a_0 a_1{-}1}
  e^{2\p i\big( {(\s_0 +1)(\d_0 a_1 -\d_1) + (\s_1 +1)(\d_1 a_0 -\d_0)
                  \over a_0 a_1 -1 } \big)}~
  s^{-\big({(\s_0+1)(a_1-1)+(\s_1+1)(a_0-1) \over a_0 a_1{-}1 }\big)}\cr
 &\qquad\quad\times
  \left[1{-}e^{2\p i\big({(\s_0+1)a_1-(\s_1+1) \over a_0 a_1{-}1}\big)}
                                                                \right]
  \left[1{-}e^{2\p i\big(
                {(\s_1+1)a_0-(\s_0+1) \over a_0 a_1{-}1}\big)}\right]\cr
 &\qquad\quad\times
  \G\Big({(\s_0+1)a_1-(\s_1+1) \over a_0 a_1{-}1}\Big)
  \G\Big({(\s_1+1)a_0-(\s_0+1) \over a_0 a_1{-}1}\Big)~.\cr}}}
 %

For longer loops with $n$ fields in the loop,
the corresponding system of $n$ equations for the $n$ $\a_i$'s
is solved as easily and the integral~\eloopper\ generalizes
straightforwardly.

\subsec{On Non-Invertible Models}\subseclab\sNonInv\noindent
The above simple models may be used as `building blocks' for \CY\ or \LGO{}
models where the defining polynomial (superpotential) is transverse and has
as many terms as there are coordinates. This by no means exhausts all the
possibilities; the recently compiled complete lists of non-degenerate
$(2,2)$ Landau-Ginzburg vacua with $c=9$~\refs{\rMaxSkII,\rMaxSkI,\rAR}
include `non-invertible' models. These are non-degenerate only upon
including more terms than there are coordinates. In addition, we may want
to consider more general theories, regardless of stringy application. Thus,
rather than trying to discuss all possible polynomials, we give one such
non-invertible example and hope that generalizations will be equally clear.

Similar to tadpoles are polynomials of the form
\eqn\eRoach{ P_0^{(R)}~ = ~x_0^{a_0} x_1 + x_1^{a_1} + x_1 x_2^{a_2}~, }
where, for homogeneity, $a_0=(d{-}k_1)/k_0$ and $a_2=(d{-}k_1)/k_2$.
The method of choosing contours is similar.
The middle term has a \prp\ as long as $x_1$
is on the spokes $x_1 = \x_1 e^{2\p i {r\over a_1}}$,
and so in a now-familiar way we choose adjacent spokes
$r=\d_1+1$ and $n=\d_1$ to form the V-shaped contour which is
equivalent to the arc at infinity.
The positivity requirements for the first and third terms lead us to
\eqn\eXXX{ \arg( x_0^{a_0}) = \arg( x_2^{a_2})~ =
           ~-{2\p( \d_1 + {1\over 2}) \over a_1}~, }
so we restrict $x_0$ and $x_2$ to be on the spokes
\eqn\eSnT{ {\twoeqsalign{
x_0 &= \x_0\, e^{2\p i({s\over a_0}-{\d_1+{1\over2}\over a_1 a_0})}
    &\qquad s\in\ZZ~,\quad 0\leq\x_0<\infty~, \cr
x_2 &= \x_2\, e^{2\p i({t\over a_2} - {\d_1+{1\over2}\over a_1 a_2})}
    &\qquad s\in\ZZ~,\quad 0\leq\x_2<\infty~, \cr}}}
and form contours with $s_{\rm in}=\d_0+1$ and $s_{\rm out}=\d_0$,
and $t_{\rm in}=\d_2+1$ and $t_{\rm out}=\d_2$.

Note, however, that the polynomial~\eRoach\ cannot be used to construct a
\LGO, as it is degenerate: $x_0=x_2=0$ and $x_1$ arbitrary parametrize a
`flat direction', where both $P_0^{(R)}$ and $\rd P_0^{(R)}$ vanish.
This is remedied~\rArnold\ by deforming $P_0^{(R)}$ into
\eqn\eRaid{ \Tw{P}_\ve^{(R)}~ = ~P_0^{(R)} + \ve x_0^{~p} x_2^{~q}~, }
where, for homogeneity, $p$ and $q$ are related by $pk_0+qk_2=a_1k_1$. It
is easy to check that, for nonzero and however small $\ve$, the deformed
polynomial~\eRaid\ is non-degenerate. The positivity condition
in~\eLaplace\ then implies a restriction on the possible values of $s,t$:
\eqn\eXXX{ k_1(\d_1+\half)-{d-k_1\over4}~ < pk_0s + qk_2t <
           k_1(\d_1+\half)+{d-k_1\over4}~, }
which may be possible to neglect for sufficiently small $\ve$. In any case,
treating $\ve$ as another parameter $\f_\a$, we calculate the period as
described above. The limit $\ve\to0$ is then taken to provide the value of
the period for the otherwise degenerate \LGO~\eRoach.

\subsec{On Fermatization}\subseclab\sFermiLoop\noindent
Alternatively, the period integrals may always be calculated by
changing variables and transforming any given model into
(an orbifold of) a `Fermat' model.
 In addition, contours for the original model may be obtained from those
in the `fermatization' using the inverse transformation.
 In practice this approach may be preferrable, especially in the cases when
part of the polynomial is of the non-invertible type discussed in the
previous subsection.

As in Ref.~\rLS, consider `fermatizing' the loop
\eqn\efermiloop{
x_1^{a_1} x_2 + x_1 x_2^{a_2} \to y_1^{b_1} + y_2^{b_2} ~.}
For the Jacobian to be constant we need
\eqn\eBeez{ b_1 = {a_1 a_2 - 1\over a_2-1}~, \qquad
            b_2 = {a_1 a_2 - 1\over a_1-1}~, }
and hence the coordinate transformation is
\eqn\eXXX{{\twoeqsalign{
 y_1 &= x_1^{a_1(a_2-1)\over a_1a_2-1} x_2^{a_2-1\over a_1a_2-1}
 \quad&\quad
 x_1 &= y_1^{a_2 \over a_2-1} y_2^{-1\over a_1-1}                \cr
 y_2 &= x_1^{a_1-1\over a_1a_2-1} x_2^{a_2(a_1-1)\over a_1a_2-1}
 \quad&\quad
 x_2 &= y_1^{1 \over a_2-1} y_2^{a_1\over a_1-1}                 \cr}}}
The type of integral that comes up in period calculations then becomes
\eqn\eYint{{\eqalign{
&\int \rd^2 x~ x_1^{\s_1}~ x_2^{\s_2}~
 e^{ -s \left( x_1^{a_1} x_2 + x_1 x_2^{a_2}\right) } =             \cr
&\qquad{a_1 a_2{-}1\over (a_1{-}1)(a_2{-}1)}
 \int \rd^2 y\
  y_1^{\s_1 a_2 - \s_2\over a_2-1} y_2^{\s_2 a_1 - \s_1\over a_1-1}
    e^{-s\big(y_1^{a_1 a_2-1\over a_2-1} +
              y_2^{a_1 a_2-1\over a_1-1}\big)}                     \cr}}}
Calculating the second double integral \eYint\
in the manner of \SS\sBPFWZ\ gives
the same answer as calculating the original double integral \eloopper\
in the  manner of \SS\sLoops.

One may worry that the fractional exponents $b_i$ may be smaller than
2 and hence our arguments used for the Fermat models of \SS\sBPFWZ\
would not be relevant. However, this is not the case.
To see this in the situation described above, rewrite Eqs.~\eBeez\ as
\eqn\eBEEZ{ b_1 = \Big({a_1-1\over a_2-1}\Big) + a_1~, \quad
            b_2 = \Big({a_2-1\over a_1-1}\Big) + a_2~, \qquad
            a_1,a_2 \in \ZZ~. }
Noting that $a_i\ge 2$ we find that this is the case for the $b_i$ as
well. Thus our results from \SS\sBPFWZ\ still holds. This argument
can be seen to apply for any other `fermatization' as well.
 Finally, if $a_1\ne a_2$, at least one of $b_i$ is non-integral.
While this does not hinder the evaluation of the period integral, a Fermat
model with integral powers is easily obtained through a further change of
variables, compensated by passing to a suitable quotient.

\subsec{Landau-Ginzburg Vacua}
\subseclab\sLGV\noindent
In the previous section we discussed a rather large class of \CY\ models,
not necessarily of dimension three, for which we can write down a set of
periods in a manifest homology basis. It is, however, clear that we do not
need to restrict to theories for which a nonlinear \CY\ $\s$-model
interpretation is known. Such examples are found among Landau-Ginzburg
string vacua with more than five superfields; for a  recent classification
of such theories see \rAR.

Let us consider the usual action of an $N=2$ superconformal
Landau-Ginzburg theory
\eqn\eACTION{\int d^2z\, d^2\q\, d^2\bar\q\, K(X_i,\bar X_i) +
             (\int d^2z\, d^2\q\, P(X_i) + c.c).}
where $K$ is the K\"ahler potential and $P$, the superpotential, is a
holomorphic function of the $N=2$ chiral superfields
$X_i(z,\bar z,\q^+,\q^-)$. Due to
nonrenormalization theorems, $P$ is not renormalized (up to scaling)
and hence will
characterize the theory (modulo irrelevant perturbations coming from
the K\"ahler potential \rVW.) Let $P$ be a polynomial in the
superfields $X_i~,~i=0,\ldots,n$, and, moreover, let it be
quasi-homogeneous of degree $d$,
 \ie, under rescaling of the world-sheet
\eqn\eSS{ X_i \mapsto \l^{k_i} X_i~~,\qquad P(X_i) \mapsto \l^d P(X_i)~.}
 The central charge is straightforward to compute \rVW ,
and is given by $c=6\sum_{i=0}^n
({1\over 2} - q_i)$ with $q_i=k_i/d$ the charge under the
left-moving $U(1)$-current $J_0$.
In order to ensure $(2,2)$ world-sheet supersymmetry we need to consider
the \LGO{} $P(X_i)/j$ where $j=e^{2\p i J_0}$. This projects
onto the integer charged states and preserves the supersymmetry.
We will not restrict our attention to models with $c=9$. Rather,
set $\hat c=c/3$ and the expression for the central charge can
rewritten as
\eqn\eCCC{
\sum_{i=0}^n k_i = k\,d
}
where $k=(n+1-\hat c)/2$.

For our purposes it is sufficient to note the isomorphism between the
chiral ring  and the ring of polynomials modulo the ideal generated by
$\vd P(x_i)$; the $x_i$ are the lowest order componets of the chiral
superfields  $X_i$ in the point field limit.

Following our discussion in section~2 the  relevant object to consider
for the periods is
given by
\eqn\eOMG{
\w_j(\f_\a)~=~C\int_{\G_j}{\prod_{i=0}^n \rd\,x_i\over (P_\f)^k}~;
}
for $n+1-\hat c=2$ this reproduces~\eOM. In particular
$n=4$, $\hat c=3$ we get the well-known expression for the period
of a \CY\ three-fold. Note
that because of  the scale invariance the integrand only depends on $n$ of
the $n+1$  coordinates. Thus, just as for the \CY\ models we can choose to
set one of the $x_i$ to one and we obtain the equivalent of~\eOM. The
polynomial is then expanded about the $\f_\a$ by splitting $P_\f$ into a
reference polynomial $P_0$ and a perturbation $\D$; thus,
\eqn\eLaplace{
\eqalign{{1\over P^k_\f} &=
{1\over (P_0 - \D)^k} = \sum_{n=0}^\infty
{(n+k-1)!\over n!}{\D^n \over P_0^{n+k} }\cr
&= \sum_{n=0}^\infty {\D^n\over n!} \int_0^\infty ds s^{n+k-1} e^{-s P_0}
{}~,\quad \Ree(P_0)>0~.}}
 The discussion for the various types of polynomials then applies
{\it verbatim} and we refer the reader to the following section
for an example and explicit calculation.

\newsec{Examples}\seclab\sExamples\noindent
Having found suitable contours for the building blocks of
reference polynomials, we may apply our results to
finding periods for any of the types listed in \rCfC.
Further, this method adapts well to many
(weighted) complete
intersection spaces~\refs{\rBeast, \rCYCI, \rGH, \rCDLS, \rGRY},
generalized Calabi-Yau manifolds~\refs{\rZ},
and Landau--Ginzburg vacua~\refs{\rMaxSkII,\rMaxSkI,\rAR}.

\subsec{A Known Example}\subseclab\sKnown\noindent
We start with a 2-parameter family of hypersurfaces
 $\CP4_{(3,2,2,7,7)}[21]^{50,11}_{-78}$ from Ref.~\rGoryVII:
\eqn\eXXX{ P_{\f_0,\f_1}~= ~x_0^7 + x_1^7 x_3 +  x_3^3 + x_2^7 x_4 + x_4^3
                         - \f_0\, x_0 x_1 x_2 x_3 x_4
                         + \f_1\, x_0^3 x_1^3 x_2^3~; }
the superscripts $50,11$ indicate the Hodge numbers $b_{2,1},b_{1,1}$ and
the subscript $+6$ is the Euler characteristic $\EU=2(b_{1,1}-b_{2,1})$.
We will calculate the periods in the coordinate patch where $x_0=1$,
and use the general results of \SS\sTadpoles.
\eqn\eKnownper{
{\eqalign{ \vp_{(\d_1,\d_2,\d_3,\d_4)}
 &= C\,\int_{(\d_1,\d_2,\d_3,\d_4)} \rd^4 x~ \int_0^\infty \rd s~
    e^{ -s P |_{x_0=1} }\cr
 &= C\,\sum_{n=0}^\infty\sum_{m=0}^\infty
       {\f_0^n\over n!}{(-\f_1)^m\over m!}~
         \int_0^\infty \rd s\ e^{-s} s^{n+m}                      \cr
 &\qquad\quad\times
    \int_{(\d_1,\d_2,\d_3,\d_4)} \rd^4 x \
    e^{-s( x_1^7 x_3 + x_3^3 + x_2^7 x_4 + x_4^3)}\
    x_1^{n+3m} \, x_2^{n+3m}\,x_3^n\,x_4^n \cr
  &={C\over 3^2 7^2}\,\sum_{n=0}^\infty\sum_{m=0}^\infty
    {\f_0^n\over n!}{(-\f_1)^m\over m!} \
    e^{2\p i\left( (\d_1+\d_2)({n+3m+1\over 7}) +
                    (\d_3+\d_4)({2n-m+2\over 7})  \right) }       \cr
  &\qquad\quad\times
    \left(1{-}e^{2\p i ({n+3m+1\over 7})}\right)^2~
     \left(1{-}e^{2\p i ({2n-m+2\over 7})}\right)^2                 \cr
  &\qquad\qquad\times
    \G^3\Big({n+3m+1\over 7}\Big) \G^2\Big({2n-m+2\over 7}\Big)~, \cr}}}
where the range of the $\d_i$ are  $0\leq\d_1,\d_2\leq 5$,
$0\leq\d_3,\d_4\leq 1$. This would lead to a set of $2^2 6^2=144$ possible
contours, but only six are in fact independent. This is easily seen as
follows.
For the deformations considered, the  period dependence on $\d_1$ is the
same as that on $\d_2$,  and the dependence on $\d_3$ and $\d_4$ is just
twice that of the dependence on $\d_1$.
That is, letting $\d_3\to\d_3+1$ is the same as letting $\d_1\to\d_1+2$.
Knowing this, we may choose $\d_2=\d_3=\d_4=0$.

Rearranging the expression for the periods by using
\eqn\eInversion{
  \left( 1 - e^{2\p i \a}\right) \G(\a) =
  - 2\p i { e^{\p i \a} \over \G(1-\a)} ~,}
and writing $l = n+1$, we arrive at the more compact expression for the
periods
\eqn\eKnownperII{
\vp_{(\d_1)}^{(0)} =
   {(2\p)^4\, C\over 3^2 7^2\,\f_0}\,
   \sum_{l=1}^\infty\sum_{m=0}^\infty
    {e^{2\p i l({\d_1+3\over 7})}\f_0^l\over \G(l)}\,
    {e^{\p i m ({6\d_1-3\over 7})}\f_1^m\over m!} \
    { \G\big({l+3m\over 7}\big) \over
      \G^2\big(1-{l+3m\over 7}\big)\G^2\big(1-{2l-m\over 7}\big)}~. }
For $\d_1=0$, this is the period calculated in Ref.~\rGoryVII.

The same example, when calculated in the patch $x_3=1$, yields
almost the same periods. After all the rearrangements and upon realizing
that the different periods can all be obtained by varying only $\d_0$,
we get
\eqn\XXX{
\vp_{(\d_0)}^{(3)} =
   {(2\p)^4\, C\over 3^2 7^2\,\f_0}\,
   \sum_{l=1}^\infty\sum_{m=0}^\infty
    {e^{2\p i l({\d_0+3\over 7})}\f_0^l~
     e^{\p i m ({6\d_0-3\over 7})}\f_1^m\over
      \big(1+e^{2\p i({l+3m\over 7})} \big) \G(l)  m!}~
    { \G\big({l+3m\over 7}\big) \over
      \G^2\big(1-{l+3m\over 7}\big)\G^2\big(1-{2l-m\over 7}\big)}~. }
We find that
\eqn\eXXX{  \vp_{(\d_0)}^{(3)} + \vp_{(\d_0+1)}^{(3)}~
          = ~\vp_{(\d_1)}^{(0)}~. }
that is, we recover the expression in $x_0=1$ coordinate patch.  So
the periods in different patches are just linear combinations of each
other.

Most notably, the action of (a subgroup of) the modular group found in
Ref.~\rGoryVII, can be realized in the present analysis simply by changing
the choice of the poly-contour:
\eqn\eModG{ (\f_0,\f_1) \longrightarrow (\l\f_0,\l^3\f_1)
            \qquad\cong\qquad \d_1 \longrightarrow \d_1+1~. }
The erudite Reader will be reminded of Dehn twists, which in the case of
Riemann surfaces generate the modular group; however, for the case of \CY\
3-spaces, the general theory of moduli space and modular group is far from
that well developed and we do not know how much of the modular group can be
uncovered in this simple fashion. We hope to return to a more detailed
analysis of this relation between these simple operations on the contours
and periods, and the elements of the modular group.

\subsec{A Mixed Example}\subseclab\sLurp\noindent
Consider the somewhat peculiar example found as the penultimate item in
Table~2 of Ref.~\rBH. It concerns the family of models
 $\IP_{(2,17,17,6,9)}[51]^{31,34}_{+6}$ and their mirror models,
 $\big\{\IP_{(3,16,17,6,9)}[51]/\ZZ_2:(0,0,0,1,1)\big\}^{34,31}_{-6}$.
The reference defining polynomials (superpotentials) are
\eqn\eMirDP{{\eqalign{
      P_0 &= \big[ x_0^{~17}x_1 + x_1^{~3} \big]~ + ~x_2^{~3}~ +
             \big[ x_3^{~7}x_4 + x_3 x_4^{~5} \big]~,              \cr
 \ha{P}_0 &= \big[ y_0^{~17} + y_0 y_1^{~3} \big]~ + ~y_2^{~3}~ +
             \big[ y_3^{~7}y_4 + y_3 y_4^{~5} \big]~.              \cr}}}
The square brackets merely group the irreducible models: a `tadpole', a
`Fermat' and a `loop' term; $\ha{P}_0$ was obtained by
`transposition'~\rBH.

Suppose we add the fundamental deformation to $P_0$, so that
$P = P_0 - \f_0 x_0 x_1 x_2 x_3 x_4$.
Then the periods in the coordinate patch where $x_2=1$ become
\eqn\eXXX{{\eqalign{
 \vp^{(2)} &=
 {C \over 6\cdot 17^2} \sum_{m=0}^\infty
  e^{2 \pi i(m+1)\left({1\over 17}\d_0+{16\over 51}\d_1+
    {2\over 17}\d_3+{3\over 17}\d_4\right)} {\f_0^m \over m!}\cr
&\qquad\times
 \left(1{-}e^{2 \pi i\left({m+1\over 17}\right)}\right)
 \left(1{-}e^{2 \pi i\left({16(m+1)\over 51}\right)}\right)
 \left(1{-}e^{2 \pi i\left({2(m+1)\over 17}\right)}\right)
 \left(1{-}e^{2 \pi i\left({3(m+1)\over 17}\right)}\right)\cr
&\qquad\times
 \G\left({m+1\over 17}\right)
 \G\left({16(m+1)\over 51}\right)
 \G\left({2(m+1)\over 17}\right)
 \G\left({3(m+1)\over 17}\right)
 \G\left({m+1\over 3}\right)             ~.\cr} }}
We see that the dependence on different contours can be summed up
by varying only $\d_1$, and so, with a slight change of notation, we find
\eqn\ePerMixA{{\eqalign{
 \vp^{(2)}_{~\d} &=
 {C \over 6\cdot 17^2} \sum_{m=0}^\infty
  e^{2 \pi i\d\left({(m+1)\over 51}\right)} {\f_0^m \over m!}\cr
&\qquad\times
 \left(1{-}e^{2 \pi i\left({m+1\over 17}\right)}\right)
 \left(1{-}e^{2 \pi i\left({16(m+1)\over 51}\right)}\right)
 \left(1{-}e^{2 \pi i\left({2(m+1)\over 17}\right)}\right)
 \left(1{-}e^{2 \pi i\left({3(m+1)\over 17}\right)}\right)\cr
&\qquad\times
 \G\left({m+1\over 17}\right)
 \G\left({16(m+1)\over 51}\right)
 \G\left({2(m+1)\over 17}\right)
 \G\left({3(m+1)\over 17}\right)
 \G\left({m+1\over 3}\right)         ~.\cr} }}

Now if we add another deformation so that
$P=P_0 - \f_0 x_0 x_1 x_2 x_3 x_4 -\f_1 x_0^{17} x_2$,
we discover that the periods are
\eqn\ePerMixB{{\eqalign{
 \vp^{(2)} &=
 {C \over 6\cdot 17^2} \sum_{m=0}^\infty \sum_{n=0}^\infty
  e^{2\pi i\left( {m+1\over 17}\d_0 +{16m-17n+16\over 51}\d_1+
                {2(m+1)\over 17}\d_3 +{3(m+1)\over 17}\d_4 \right) }
{\f_0^m \over m!}{\f_1^n \over n!}\cr
&\times
 \left(1{-}e^{2 \pi i\left({m+1\over 17}\right)}\right)
 \left(1{-}e^{2 \pi i\left({16m-17n+16\over 51}\right)}\right)
 \left(1{-}e^{2 \pi i\left({2(m+1)\over 17}\right)}\right)
 \left(1{-}e^{2 \pi i\left({3(m+1)\over 17}\right)}\right)\cr
&\times
 \G\left({m+17n+1\over 17}\right)
 \G\left({16m-17n+16\over 51}\right)
 \G\left({2(m+1)\over 17}\right)
 \G\left({3(m+1)\over 17}\right)\cr
&\times
 \G\left({m+n+1\over 3}\right) ~.\cr} }}
We need only vary $\d_0$ and $\d_1$ to find all the periods.

Adding one more deformation, the polynomial becomes
$P=P_0 - \f_0 x_0 x_1 x_2 x_3 x_4 -\f_1 x_0^{17} x_2-\f_2 x_3^4 x_4^3$.
The periods are then easily seen to be:
\eqn\ePerMixC{{\eqalign{
 \vp^{(2)} &=
 {C \over 6\cdot 17^2} \sum_{m=0}^\infty \sum_{n=0}^\infty\sum_{p=0}^\infty
  e^{2\pi i\left( {m+1\over 17}\d_0 +{16m-17n+16\over 51}\d_1+
                {4m+17p+4\over 34}\d_3 +{6m+17p+6\over 34}\d_4 \right) }
{\f_0^m \over m!}{\f_1^n \over n!}{\f_2^p \over p!}\cr
&\times
 \left(1{-}e^{2 \pi i\left({m+1\over 17}\right)}\right)
 \left(1{-}e^{2 \pi i\left({16m-17n+16\over 51}\right)}\right)
 \left(1{-}e^{2 \pi i\left({4m+17p+4\over 34}\right)}\right)
 \left(1{-}e^{2 \pi i\left({6m+17p+6\over 34}\right)}\right) \cr
&\times
 \G\left({m+17n+1\over 17}\right)
 \G\left({16m-17n+16\over 51}\right)
 \G\left({4m+17p+4\over 34}\right)
 \G\left({6m+17p+6\over 34}\right)\cr
&\times
 \G\left({m+n+1\over 3}\right)~.\cr} }}

Judicious use of Eq.~\eInversion \ allows the periods to be
written in a more compact form;
we display this series of sets of periods in this fashion
to show the ease of application of our formalism.
The Reader will see that adding further perturbations
presents no barrier to the calculation of periods.

\subsec{A Twisted Example}\subseclab\sFract\noindent
The first example in Ref.~\rLS\ concerns a pair of manifolds
related by a fractional transformation
\eqn\eXXX{{
\vbox{
\tabskip=25pt\halign{%
\hfil$#$:\hfil&$#$\hfil&$#$\hfil\cr
{\cal M} & \IP_{(1,7,2,2,2)}[14]^{122,2}_{-240} &
      P_0 = x_1^{14} + x_2^2 + x_3^7 + x_4^7 + x_5^7 \cropen{8pt}
\tilde{\cal M} & \IP_{(1,3,1,1,1)}[7]^{122,2}_{-240} &
      \tilde{P_0} = y_1^{7} + y_1 y_2^2 + y_3^7 + y_4^7 + y_5^7 ~.\cr} }}}
All 122 $b_{21}$ states of $\tilde{\cal M}$ are represented
by deformations of the polynomial $\tilde{P_0}$,
and are untwisted states in the corresponding Landau-Ginzburg model.
{}From the fractional transformation
\eqn\eFT{{
\vbox{\tabskip=25pt\halign{%
$#$\hfil&$#$\hfil\cr
y_1 = x_1^2 & x_1 = y_1^{{1\over 2}} \cropen{8pt}
y_2 = x_2/x_1 & x_2 = y_1^{{1\over 2}} y_2 \cr}} }}
we see that there are 15 deformations
\eqn\eYstates{{
\left\{ y_2 \right\}\otimes
\{y_3^{a_3}y_4^{a_4}y_5^{a_5} ~|~0\leq a_i \leq 4,~a_3+a_4+a_5=4\}}}
which get mapped to rational functions
\eqn\eYimage{{
\left\{ {x_2\over x_1} \right\}\otimes
\{x_3^{a_3}x_4^{a_4}x_5^{a_5} ~|~0\leq a_i \leq 4,~a_3+a_4+a_5=4\}~.}}
Analyzing the Landau-Ginzburg model described by $P_0$
we find 107 untwisted and 15 twisted states,
the former images under \eFT\ of deformations of $\tilde P_0$
and the latter of the form
\eqn\eXstates{{
\left\{ \vert (\frc{3}{7},\frc{3}{7}\rangle_{(c,c)}^7 \right\}\otimes
\{x_3^{a_3}x_4^{a_4}x_5^{a_5} ~|~0\leq a_i \leq 4,~a_3+a_4+a_5=4\}~.}}
Thus we conclude that there is a correspondence between the twisted vacuum
and a rational function,
${x_2\over x_1} \leftrightarrow
\vert (\frc{3}{7},\frc{3}{7}\rangle_{(c,c)}^7.$
Note that there is no deformation of $P_0$
that is not mapped to a deformation of $\tilde{P_0}$.
One might have thought from \eFT\ that some square roots of
$y_1$ might pop up, but a simple analysis shows that they cannot.
To make a deformation of $P_0$ of charge 1,
every odd power of $x_1$ must be accompanied
by an odd power of $x_2$, and {\it vice versa},
and hence the image under \eFT\
has an integral power of $y_1$.  This is related to the fact that the
$\ZZ_2$ identification necessary to make \eFT\ a one-one map
is part of the projective equivalence identification on ${\cal M}$.
The $\ZZ_2$ acts on the $x$'s as
\eqn\eZtwo{
 (x_1,x_2,x_3,x_4,x_5) \rightarrow (\a x_1,\a x_2,x_3,x_4,x_5)}
where $\a^2=1$, and thus the $\ZZ_2$
is seen to be part of the projective equivalence identification,
$$\IP_{(1,7,2,2,2)} = \IC_5/\sim  ~,$$
with $(x_1,x_2,x_3,x_4,x_5) \sim (\l x_1,\l^7 x_2,\l^2 x_3,\l^2 x_4,\l^2 x_5)$,
where $\l\in\IC^{\star}$.

Suppose we add the fundamental deformation and one more to $P_0$, so that
$P = P_0 - \f_0 x_1 x_2 x_3 x_4 x_5 - \f_1 x_1^6 x_3^2 x_4^2$.
Then the periods in the coordinate patch where $x_5=1$ are
\eqn\ePerFTx{{\eqalign{
 \vp^{(5)} &=
 {C_x \over 2^2\cdot 7^3} \sum_{m=0}^\infty \sum_{n=0}^\infty
 e^{2\pi i\left( {m+6n+1\over 14}\d_1 +{m+2n+1\over 7}(\d_3+\d_4)\right) }
 {\f_0^m \over m!}{\f_1^n \over n!}\cr
&\times
 \left(1{-}e^{2 \pi i\left({m+6n+1\over 14}\right)}\right)
 \left(1{-}e^{2 \pi i\left({m+1\over 2}\right)}\right)
 \left(1{-}e^{2 \pi i\left({m+2n+1\over 7}\right)}\right)^2\cr
&\times
 \G\left({m+6n+1\over 14}\right)
 \G\left({m+1\over 2}\right)
 \G^2\left({m+2n+1\over 7}\right)
 \G\left({m+1\over 7}\right)~.\cr} }}
We need only vary $\d_1$ and $\d_3$ to find all the periods.

Suppose we add the images of these deformations under
\eFT\ to $\tilde P_0$, so that
$\tilde P = \tilde P_0 -
\psi_0 y_1 y_2 y_3 y_4 y_5 - \psi_1 y_1^3 y_3^2 y_4^2$.
Then the periods in the coordinate patch where $y_5=1$ are
\eqn\ePerFTy{{\eqalign{
 \tilde\vp^{(5)} &=
 {C_y \over 2\cdot 7^3} \sum_{m=0}^\infty \sum_{n=0}^\infty
 e^{2\pi i\left( {m+6n+1\over 14}\d_1 +{m+2n+1\over 7}(\d_3+\d_4)\right) }
 {\psi_0^m \over m!}{\psi_1^n \over n!}\cr
&\times
 \left(1{-}e^{2 \pi i\left({m+6n+1\over 14}\right)}\right)
 \left(1{-}e^{2 \pi i\left({m+1\over 2}\right)}\right)
 \left(1{-}e^{2 \pi i\left({m+2n+1\over 7}\right)}\right)^2\cr
&\times
 \G\left({m+6n+1\over 14}\right)
 \G\left({m+1\over 2}\right)
 \G^2\left({m+2n+1\over 7}\right)
 \G\left({m+1\over 7}\right)~.\cr} }}
These are the same as \ePerFTx\ up to a normalization factor,
letting $\f_i=\psi_i$.
This is not surprising - the period calculations
are related simply by the change of variables~\eFT.

The value of this comes when we want to look at the dependence of
the periods of ${\cal M}$ on twisted moduli.
This we can do by looking at the dependence of the periods of
$\tilde{\cal M}$ on the deformations \eYstates.
So suppose
$\tilde P = \tilde P_0 -
\psi_0 y_1 y_2 y_3 y_4 y_5 - \psi_2 y_2 y_3^4 $.
Then the periods in the coordinate patch where $y_5=1$ are
\eqn\ePerFTytwist{{\eqalign{
 \tilde\vp^{(5)} &=
 {C_y \over 2\cdot 7^3} \sum_{m=0}^\infty \sum_{n=0}^\infty
 e^{2\pi i\left( {m-n+1\over 14}\d_1 +{m+4n+1\over 7}\d_3+
                {m+1\over 7}\d_4\right) }
 {\psi_0^m \over m!}{\psi_2^n \over n!}\cr
&\times
 \left(1{-}e^{2 \pi i\left({m+n+1\over 2}\right)}\right)
 \left(1{-}e^{2 \pi i\left({m-n+1\over 14}\right)}\right)
 \left(1{-}e^{2 \pi i\left({m+4n+1\over 7}\right)}\right)
 \left(1{-}e^{2 \pi i\left({m+1\over 7}\right)}\right)\cr
&\times
 \G\left({m+n+1\over 2}\right)
 \G\left({m-n+1\over 14}\right)
 \G\left({m+4n+1\over 7}\right)
 \G^2\left({m+1\over 7}\right)~.\cr} }}
Reinterpreted as periods of ${\cal M}$,
the $\psi_2$ dependence is the dependence
on the modulus associated to a twisted state, \ie,
one of the states in \eXstates.
Note that the same calculation could have been done by adding
the deformations in \eYimage\ to $\tilde P_0$.

\subsec{A Higher Dimensional Example}

Let us consider the following example: $\IC^7_{1,6,6,4,4,6,9}[18]$,
the zero locus of a defining polynomial of degree 18.
Take the polynomial to be
\eqn\ehdw{
P_\f~=~x_0^{18} + x_1^3 + x_2^3 + x_1 x_3^3 + x_2 x_4^3 + x_5^3 + x_6^2
-\f_0 x_0^4 x_3x_4x_5 -\f_1 x_0^6 x_1x_2~.}
Following the results of section~2 and the discussion above
it is now straightforward to
write down the periods associated to the above family.
In patch ${\cal U}_0$ we have
\eqn\ehdper{
{\eqalign{\vp_{(\d_1,\ldots,\d_6)}&=C\int_{(\d_1,\ldots,\d_6)}
{\prod_{i=1}^6 \rd\,x_i \over (P_\f |_{x_0=1})^2} \cr
&=C\,\sum_{n,m=0}^\infty
       {\f_0^n\over n!}{\f_1^m\over m!}~
         \int_0^\infty \rd s\ e^{-s} s^{n+m+1}\cr
 &\qquad\times
    \int_{(\d_1,\ldots,\d_6)} \rd^6 x
    e^{-s( x_1^3 +x_1x_3^3 + x_2^3 + x_2 x_4^3 + x_5^3 +x_6^2)}\
    x_1^m \, x_2^m\,x_3^n\,x_4^n\, x_5^n \cropen{5pt}
{}~&=~C\,{\G(1/2)\over 3^5}\sum_{n,m=0}^\infty {\f_0^n\over n!}{\f_1^m\over
m!} e^{2\p i\left((\sum_{i=3}^5\d_i)({n+1\over 3})+
(\d_1+\d_2)({3(m+1)-(n+1)\over 9})\right)} \cr
&\qquad\times~
\left(1{-}e^{2\p i({n+1\over 3})}\right)^3
\left(1{-}e^{2\p i({3(m+1)-(n+1)\over 9})}\right)^2 \cr
&\qquad\times~
\G^3\left({n+1\over3}\right)\G^2\left({3(m+1)-(n+1)\over 9}\right)
\G\left({6m+4n+1\over 18}\right)~.}}
}
Note that the period depends
only on combinations $\d_3+\d_4+\d_5$, $\d_1+\d_2$
and that letting $\d_3\to\d_3+1$ is equivalent to $\d_1\to\d_1+6$.
Thus we may set $\d_2=\ldots=\d_5=0$
if we allow $\d_1$ to take
values $0,1,\ldots,8$.
By rearranging the periods
using~\eInversion \ we finally obtain:
\eqn\ehdperII{
{\eqalign{
\vp_{(\d_1)}^{(0)}=C\,{\G(1/2)(2\p i)^5\over3^5\f_0\f_1}
\sum_{m,n=1}^\infty{e^{2\p im({\d_1+1\over3})}
e^{2\p in({7-2\d_1\over 18})}\f_0^n \f_1^m \G({6m+4n-9\over 18})
\over\G(n)\G(m) \G^3(1-{n\over3})\G^2(1-{3m-n\over9})}~.}}}
The $\d_1$ would generate nine periods but they are not all independent.
Rather, because of the factor $\G^3(1-{n\over3})$ in the denominator in
$\vp_{(\d_1)}^{(0)}$ there are the following relations:
\eqn\erelper{
\sum_{j=0,1,2}\vp_{3j+k}=0\,,\qquad k=0,1,2~.}
Thus we are left with six linearly independent periods as expected, since
we are considering a model with two complex structure deformations.

 %
\vfill\noindent
{\bf Acknowledgments}:\\
The authors acknowledge fruitful discussions with P.~Candelas and
X.~de~la~Ossa.  E.~D. would like to thank R.~Schimmrigk for discussions
of fractional transformations.
 P.~B. was supported by the American-Scandinavian Foundation, the Fulbright
Program, NSF grants PHY 8904035 and PHY 9009850, DOE grant DE-FG02-90ER4052
and the Robert~A.~Welch Foundation. P.~B. would also like to thank the ITP,
Santa Barbara and the Theory Division, CERN for their hospitality during the
initial stages of this project.
 T.H.\ was supported by the Howard University '93 FRSG Program.
E.~D. and D.~J. were supported by NSF grant PHY 9009850 and the
Robert~A.~Welch Foundation.

 %
\appendix{A}{Turning Spokes into Cycles}\noindent
 Consider a Fermat-type polynomial of weight $d$,
and suppose the weight of coordinate $x_1$ is 1.
Take a polynomial
\eqn\eXXX{{
p = x_1^d +x_2^{a_2} + x_3^{a_3} + x_4^{a_4} + x_5^{a_5}
      -\psi x_1 x_2 x_3 x_4 x_5~. }}
Now the symmetries of this polynomial include the following, listed
along with their actions\penalty10000\ft{We will use
the notation $(\ZZ_k:\Q_1,\ldots,\Q_5)$ for a $\ZZ_k$
symmetry with the action\hfill\break $(X_1,\ldots,X_5) \to
(\a^{\Q_1} X_1,\ldots,\a^{\Q_5} X_5)$, where
$\a^k~=~1$.}:
\eqn\eSym{{\eqalign{
 g_1&= (\ZZ_d ~:~ 1,k_2,k_3,k_4,k_5)~, \cr
 g_2&=(\ZZ_{a_2} ~:~ a_2{-}1,1,0,0,0)~, \cr
 g_3&=(\ZZ_{a_3} ~:~ a_3{-}1,0,1,0,0)~, \cr
 g_4&=(\ZZ_{a_4} ~:~ a_4{-}1,0,0,1,0)~, \cr
 g_5&=(\ZZ_{a_5} ~:~ a_5{-}1,0,0,0,1)~. \cr} }}
The inhomogeneous coordinates in the patch ${\cal U}_l$ ($l\neq 1$)
are
\eqn\eAfFine{ \x^{(l)}_i~ \define ~\Big(
               {x_i \over x_l^{k_i/k_l}}\Big)
              ~~{\rm in}~~{\cal U}_l,~~{\rm where}~~x_l\ne0~. }
The building blocks for the cycles are three-chains,
written in any patch
${\cal U}_l$ ($l\neq 1$) as
\eqn\eChain{\eqalign{
  V^{(l)}_j = \{ &\x^{(l)}_k | \x^{(l)}_l=1;
  \x^{(l)}_i\ {\rm real\ and\ positive,\ for\ }i\neq 1,l; \cr
    &\x^{(l)}_1\  {\rm is\ a\ solution\ of}\  p=0
  \ {\rm on\ the\ branch}\ \arg(\x^{(l)}_1) \rightarrow
   \pi +{2 \pi j \over d}\ {\rm as}\ \psi\rightarrow 0 \}~.\cr}}
The three-chain on the manifold is then $V_j=\cup V^{(l)}_j$.

In a patch ${\cal U}_l$ there are 3 two-dimensional boundaries
to $V^{(l)}_j$,
which we denote $B^{(l)}_{j,i}\  , \quad i=2,3,4$.
These occur where one of the $\x^{(l)}_i$ vanishes
and the other two do not.
On the overlap of two patches,
\eqn\eXXX{B^{(l)}_{j,i} = B^{(m)}_{j,i}\quad{\rm if}\quad i\neq l\neq m~.}
Note that $B^{(l)}_{j,i}$ is not in the patch ${\cal U}_i.$
Over the manifold, then, there are four independent components
to the boundary of $V_j$:
\eqn\eFoo{\eqalign{
B_{j,2} &= B^{(3)}_{j,2} = B^{(4)}_{j,2} = B^{(5)}_{j,2}~, \cr
B_{j,3} &= B^{(2)}_{j,3} = B^{(4)}_{j,3} = B^{(5)}_{j,3}~, \cr
B_{j,4} &= B^{(2)}_{j,4} = B^{(3)}_{j,4} = B^{(5)}_{j,4}~, \cr
B_{j,5} &= B^{(2)}_{j,5} = B^{(3)}_{j,5} = B^{(4)}_{j,5}~; \cr}}
only three will be manifest in any patch.

Let us restrict to patch ${\cal U}_5$,
and momentarily drop the superscript labelling the patch.
Now if $A$ is the operation that takes
$\psi\rightarrow e^{-{2 \pi i \over d}} \psi$, then
certain powers of $A$ are equivalent to a coordinate transformation
\eqn\eAAA{{
A^{k_j} : x_j \rightarrow e^{2 \pi i k_j\over d} \, x_j ~.}}
The integration contours using adjacent spokes in
$\x_2$, $\x_3$, and $\x_4$ are therefore formed by
\eqn\eXXX{{
Q = \left( 1-A^{k_2} \right)\left( 1-A^{k_3} \right)
    \left( 1-A^{k_4} \right) V_j ~.}}
Note that $Q$ has no boundary of the form
$B_{m,4}$
owing to the fact that $\left(1-A^{k_4}\right)B_{j,4} = 0$.
Similarly, all the $B_{m,2}$ and $B_{m,3}$ components vanish,
so $Q$ is indeed a cycle.

 In a quotient by~\eSym\ (say, for construction of the mirror model), the
boundaries $B_{m,2}$'s become identified, and similarly the $B_{m,3}$'s,
and so on. Then we may write
\eqn\eXXX{{\eqalign{
Q &= \left( 1-A^{k_2} \right)\left( 1-A^{k_3} \right)
\left( 1-A^{k_4} \right) V_j \cr
  &= V_j - V_{j+k_2} - V_{j+k_3}- V_{j+k_4} + V_{j+k_2+ k_3}
     + V_{j+k_2+ k_4}+ V_{j+k_3+ k_4}- V_{j+k_2+ k_3+k_4}~.}}}

A similar approach may be used to show that the
relevant contours for the other polynomial types are cycles as well.
In particular, any polynomial can be put in a Fermat form
by fermatization, as described in section~2.5.

\vfill
\ifx\answ\bigans\eject\else\fi

\listrefs

 %
\bye